\documentclass[12pt]{article}
\usepackage{amssymb,amsmath,esint,titlesec}
\titleformat*{\section}{\normalsize\bf}
\titleformat*{\subsection}{\small\bf}

\begin{document}


\begin{titlepage}

\setlength{\baselineskip}{18pt}

                               \vspace*{0mm}

                             \begin{center}

{\LARGE\bf Embolic aspects of black hole entropy}

                            \vspace*{3.5cm}

              \large\sf  NIKOLAOS  \  \  KALOGEROPOULOS $^\dagger$\\

                            \vspace{0.2cm}
                            
 \normalsize\sf Center for Research and Applications \\
                                  of Nonlinear Systems \ (CRANS),    \\
   University of Patras, \  Patras 26500, \ Greece.                        \\

                            \end{center}

                            \vspace{3.5cm}

                     \centerline{\normalsize\bf Abstract}
                     
                           \vspace{3mm}
                     
\normalsize\rm\setlength{\baselineskip}{18pt} 

We attempt to provide a mesoscopic treatment of the origin of black hole entropy in (3+1)-dimensional 
spacetimes.  We treat the case of horizons having space-like sections $\Sigma$ which are topological spheres,
following Hawking's and the Topological Censorship theorems.
We use the injectivity radius of the induced metric on $\Sigma$ to encode the linear dimensions of the elementary 
cells giving rise to such entropy. We use the topological entropy of $\Sigma$ as the fundamental quantity expressing the
complexity of $\Sigma$  on which its entropy depends. We point out the significance, in this context, of the Berger and Croke 
isoembolic inequalities.

                           \vfill

\noindent\small\sf Keywords: \  \  Riemannian geometry, Isoembolic inequalities, Black holes, Entropy, Topological censorship. \\
                                                                         
                             \vspace{5mm}

\noindent\small\sf Mathematics Subject Classification 2010: \  \  53Z05, 58Z05, 37N20, 83C57.

                             \vfill

\noindent\rule{9cm}{0.2mm}\\  
   \noindent $^\dagger$ \small\rm Electronic address: \ \  \  {\normalsize\sf   nikos.physikos@gmail.com}\\

\end{titlepage}
 

                                                                                \newpage                 

\rm\normalsize
\setlength{\baselineskip}{18pt}

\section{Introduction}

The microscopic origin of the entropy of black holes is, arguably, one of the most outstanding theoretical problems of modern day Physics \cite{Sorkin, Wald, Carlip}. 
It rests at the intersection of gravitation and quantum Physics and its ultimate resolution is widely believed to be attainable 
only when we have successfully formulated a quantum theory of gravity. Due to the fact that this goal seems to be very far off, 
despite intensive efforts since the earliest days of both General Relativity and quantum theory, we have to proceed with the conceptual and calculational
framework that  we currently have available. The opposite side of this is the belief, and hope, that exploring the microscopic origins of black hole entropy will allow 
us a glimpse into the, still elusive, regime of quantum gravity \cite{Sorkin, Wald, Carlip}.\\
  
One of the problems that someone is facing in an attempt to understand  black holes is that aspects of the stress energy tensor of physical theories are 
not very well understood. As a result, one faces substantial challenges if one wishes to quantize non-vacuum black hole solutions to the Einstein
equations. Most of the classical results in 
black hole Physics assume a uniform lower bound on the stress energy tensor \cite{HE}, in the form of classical energy conditions, which appear reasonable from the physical standpoint
and are  necessary  technically, to establish key results such as the Raychaudhuri equation for instance \cite{HE}, 
in order to prove / reach non-trivial conclusions.\\

Theoretically one cannot ``naturally" exclude states giving rise to unbounded from below expectation values of the stress enegy tensor. One way around this difficulty has been to 
generalize the classical energy conditions to hold ``on the average", instead of point-wise, in spacetime \cite{Ford, Fewster}. This approach has had some modest success. In the present work we follow a completely 
different approach. Instead of trying to get bounds on the stress energy tensor, we  ignore the possibility of such lower bounds. To do this effectively, and still maintain some connection 
to the Einstein equations, amounts to making statements about the black hole  that are independent of the Ricci tensor (or Ricci curvature) of space-time. Hence we are lead to consider curvature-free 
statements pertinent to black holes space-times, and more specificaly to their horizons. Many of such results are obtained in the contexts of systolic and embolic inequalities \cite{Berger}, which we propose to apply to space-like sections of the black hole horizons, to  properties of which the entropy of the black holes is ascribed \cite{Sorkin, Wald, Carlip}. \\         
 
To this end, the present note  should be considered independent of, but  complementary, to \cite{NK1}, where a conjecture about the mesoscopic origin of the entropy of black holes in (3+1)-dimensional space-time was 
presented.
In that work, the entropy of a black hole was ascribed to the non-trivial homotopy type of the space-like sections \ $\Sigma$ \ of its event horizon, or in a more quasi-
local 
context, to space-time's marginally outer trapped surfaces. We treated the case of horizons in stationary quasi-equilibrium in order for the concept of entropy to 
be reasonably well-defined and have a physical meaning. We assumed in \cite{NK1}  a ``maximal", in some sense, violation of Hawking's \cite{Hawk} and of the Topological Censorship 
theorems \cite{FSW}. \\

In the present work, we follow a more ``conservative/conventional" approach and assume the validity of Hawking's and the Topological Censorship theorems, 
which establish the spherical topology of \ $\Sigma = \mathbb{S}^2$. \ 
The fact that topologically \ $\Sigma = \mathbb{S}^2$ \ will be assumed everywhere in the sequel.  
We ascribe the physical entropy \  $\mathcal{S}_{BGS}$ \  to be the topological entropy \ $h_{top}(\Sigma)$ \ of \ $\Sigma$, \ rather 
than a function of the volume entropy of the induced metric \ $\mathbf{g}_\Sigma$ \ on \ $\Sigma$, \ in contrast to what was done in \cite{NK1}.  
The role of the systoles of \cite{NK1},  is played in the present purely geometric framework, by the injectivity radius \ $\mathsf{inj}(\Sigma, \mathbf{g}_\Sigma )$ \  
of  \ ${\bf g}_\Sigma$. \ The role of the systolic area \ $\sigma (\Sigma, \mathbf{g}_\Sigma)$ \  in \cite{NK1}, is played 
in the present work by the embolic constant \  $\mathsf{emb}(\Sigma, \mathbf{g}_\Sigma)$.\\

The present work maintains the hand-waving style  of the  arguments of \cite{NK1}. Moreover, it is also everywhere curvature-free, something that may be 
advantageous as not much is known about the bounds of stress-energy tensors in general (3+1)-dimensional space-times, as well as about the particular space-like 
surfaces whose to whose area we ascribe the Boltzmann/Gibbs/Shannon entropy \ $\mathcal{S}_{BGS}$. \ In Section 2, we present some statements about the 
injectivity radius and the Berger and Croke embolic inequalities. In Section 3, we present our argument why we should consider the topological entropy to be the statistical entropy of the horizon.
We wrap up our discussion in Section 4 with  pointers toward future work. \\    


\section{Embolic considerations and the black hole entropy}

We recall that the injectivity radius of a Riemannian manifold ($\mathcal{M}, \mathbf{g}$) is the supremum of all \ $r>0$ \  such that the exponential map is a 
diffeomorphism  from the ball centered at the origin \ $B(0, r) \in T_P\mathcal{M}$ \  to its image on \ $\mathcal{M}$. \ From an operational viewpojnt, the injectivity radius 
defines the largest distance  from every point of 
\ $\mathcal{M}$ \ for which the normal coordinate system is valid everywhere on it, or  equivalently, the largest distance  for which the radial geodesics around any \ 
$P\in\mathcal{M}$ \ are distance minimizing. As a result, the essential feature of all balls of radius smaller than the injectivity radius is that they are contractible.\\

 In case there is an upper bound \ $\Lambda\in\mathbb{R}$ \ on the sectional curvature \ $K_\mathcal{M}$, \ namely 
if \ $K_\mathcal{M} \leq \Lambda$ \ one  gets as a direct corollary of the Rauch comparison theorem \cite{CE} that if \ $\Lambda  \leq 0$ \  then \ 
$\mathsf{inj}(\mathcal{M}, \mathbf{g})$ \ is equal to half of the diameter of \ ($\mathcal{M}, \mathbf{g}$) \ or infinite, namely that ($\mathcal{M}, \mathbf{g}$) 
has no conjugate points. If, on the other hand, \ $\Lambda > 0$, \ then for any \ $P\in\mathcal{M}$  \ and an arc-length \ $s$ \ parametrized geodesic starting 
at \ $P$, \ then there will not be any conjugate points to \ $P$ \  along this geodesic  for any \ $s < \pi / \sqrt{\Lambda}$. \ 
Moreover, \cite{CGT} prove that if for a complete $n$-dimensional Riemannian manifold \ ($\mathcal{M}, \mathbf{g}$) \ there is  a constant \ $\Lambda \geq 0$ \ 
such that its  sectional curvature \  $K$ \ satisfies \ $|K| \ \leq \Lambda$, \ and if there  exists a point \ $P\in\mathcal{M}$ \  and a constant \ $V_0 >0$ \  such that \ 
$Vol (B(P, 1)) \ \geq \ V_0$, \ then there exists a constant \ $C(n, \Lambda, V_0) \ > 0$  \ such that 
\begin{equation}
     \mathsf{inj}_P (\mathcal{M}, \mathbf{g}) \ \geq \ C
\end{equation}
In other words, at point \ $P$ \  the non-collapsing conditions expressed through the injectivity radius and the volume are equivalent.  
The above indicate that \ $\mathsf{inj}(\mathcal{M}, \mathbf{g})$ \  may serve as a geometric substitute of the length of the systole in the 
case of \ $\mathbb{S}^2$, \ where all loops are  contractible namely \ $\pi_1(\mathbb{S}^2) = 0$. \ In general, the injectivity radius is either equal to one-half 
of the shortest periodic geodesic or equal to the shortest distance between two conjugate points of \ ($\mathcal{M}, \mathbf{g}$).\\

Although we are interested only in topological spheres of dimension 2, the following definition applies to compact Riemannian manifolds irrespective of their 
dimension. Consider a compact $n$-dimensional Riemannian manifold \ ($\mathcal{M}, \mathbf{g}$). \ Then its embolic constant is defined as
\begin{equation}  
      \mathsf{emb}(\mathcal{M}) \ = \ \inf_{\mathbf{g}} \frac{Vol (\mathcal{M}, \mathbf{g})}{[\mathsf{inj}(\mathcal{M}, \mathbf{g})]^n}
\end{equation}
where the infimum is taken over all Riemannian metrics \ $\mathbf{g}$ \ on \ $\mathcal{M}$.  \  It turns out that the embolic constant of such compact 
Riemannian manifolds is bounded from below  (the Berger isoembolic inequality)  \cite{Berger1} by 
\begin{equation}
      \mathsf{emb}(\mathcal{M}) \ \geq \ \frac{\alpha_n}{\pi^n}
\end{equation}
where \ $\alpha_n$ \ indicates the volume of the round sphere \ ($\mathbb{S}^n, \mathbf{g}_0$), \ whose injectivity radius is \ 
$\mathsf{inj}(\mathbb{S}^n, \mathbf{g}_0)  = \pi$. \ In addition \cite{Berger1}, the equality in the isoembolic inequality (3)  is attained only for 
the round spheres \ ($\mathbb{S}^n, \mathbf{g}_0$). \ Elaborating upon (3), \cite{Croke1} established the Croke isoembolic inequality 
which states that every $n$-dimensional Riemannian  manifold \ $\mathcal{M}$ \ which is not a sphere satisfies
\begin{equation}
     \mathsf{emb}(\mathcal{M}) \ > \ \mathsf{emb} (\mathbb{S}^n, \mathbf{g}_0) + c(n)
\end{equation} 
where \ $c(n) >0$ \ is a  constant depending only on \  $n$. \ An indication of the difficulty of the topic of establishing embolic inequalities and 
determining embolic constants,  is that no embolic constant is explicitly 
known for any other manifold except for the $n$-sphere \ $\mathbb{S}^n$. \ Despite this, \cite{GPW} proved that the set of embolic constants
\ $\mathsf{emb}(\mathcal{M})$ \ is discrete, when \ $\mathcal{M}$ \ goes through all the manifolds of a given dimension.\\ 

We see that the embolic constant is an analogue of the systolic area, which as was mentioned in \cite{NK1} can be considered as a basic quantity 
in the determination of the entropy of \ $\Sigma$. \ Moreover, the embolic inequalitiy (3) shares the desirable feature that it is curvature-free, like the 
isosystolic inequalities whose potential physical significance was pointed out in \cite{NK1}.  \\


\section{Entropy of horizons with sections of spherical topology}

In \cite{NK1} we chose the volume entropy \ $h_{vol}$ \ (asymptotic volume) as the basic quantity encoding the entropy of \ $\Sigma$. \
This does not seem to be a good choice in the present context, as the features that made \ $h_{vol}$ \ a desirable choice in \cite{NK1} were intimately 
related to the hyperbolicity of the universal cover of \ $\Sigma$, \ something lacking in the present case. However, considering that in \cite{NK1}
we relied on the equality of the volume \ $h_{vol}$ \  and the topological entropy \ $h_{top}$, \ for  manifolds of strictly negative sectional curvature, 
it may be worth exploring the possibility that it is the topological entropy of \ $\Sigma$ \ that may have the desirable features for \ $\Sigma = \mathbb{S}^2$. \ 
For a comprehensive treatment, proofs and  references to the original works on  topological entropy, we refer to \cite{KatHas}.\\       

The entropy of the black hole expresses the ``lack of order" or ``complexity" \cite{Suss} of that system. Probably the simplest way to quantify the 
``complexity" of a Riemannian manifold is by using the level of non-integrability of its geodesic flow. In the present case, \ $\Sigma$ \ has a spherical 
topology but can have various geometries which depend not only on the matter content of the theory but also on the particular choice of \ $\Sigma$, \  
in spacetime, namely on the observer. The topological entropy \ $h_{top} (\Sigma, \mathbf{g})$ \ encodes such an exponential complexity, something 
typical of the statistical entropy of Boltzmann/Gibbs/Shannon \ $\mathcal{S}_{BGS}$. \ This motivation can be made more precise by R. Ma\~{n}\'{e}'s formula \cite{Mane}
which states the following: let \ $\Sigma$ \ be a surface endowed with a \ $C^\infty$ \ Riemannian metric \ $\mathbf{g}$. \ Let \ $x,y \in \Sigma$ \ and let \
$t > 0$. \  Moreover, define \ $n_t(x,y)$ \ to be the number of geodesic arcs joining \ $x$ \ and \ $y$ \ having length smaller than or equal to \ $t$. \ Then 
\cite{Mane}  
\begin{equation}
   \lim_{t\rightarrow +\infty} \frac{1}{t} \  \log \int_{\Sigma \times \Sigma} n_t(x,y) \  dx \ dy \ = \ h_{top}(\Sigma, \mathbf{g})
\end{equation}
So, if  \ $h_{top} > 0$, \ then the number of geodesic arcs between any two points grows exponentially, on the average, as a function of  their length with exponent
\ $h_{top}$. \ Interestingly, one cannot really get rid of  the averaging taking place under the integral sign, as it is true for all \ $x\in\Sigma$ \ and almost all
\ $y\in \Sigma$ \ that  
\begin{equation}
    \limsup _{t\rightarrow +\infty} \  \frac{1}{t} \ \log n_t(x,y) \ \leq \ h_{top}(\Sigma, \mathbf{g})
\end{equation}
and the equality does not hold as equality even for ``typical" points \ $x, y\in \mathbb{S}^2$ \ for an open set of \ $C^\infty$ \ metrics on \ $\mathbb{S}^2$. \     
To get a sense about how randomness, usually associated with lack of order and consequently with entropy,  enters this picture, 
it may be worth comparing (5) with the asymptotic equipartition property (the Shannon-McMillan-Breiman theorem) \cite{AC}. \\    

The surface \ $\Sigma = \mathbb{S}^2$ \ admits many completely integrable metrics, such as surfaces of revolution, ellipsoids and Poisson  spheres having 
zero topological entropy. On the other hand, \cite{KW} constructed metrics with \ $h_{top} > 0$ \ by considering perturbations of
 homoclinic and heteroclinic connections on an ellipsoid with three distinct axes. Subsequently \cite{Pater} constructed real analytic convex metrics
 with positive topological entropy that arose from rigid body dynamics. Eventually, \cite{ConPat} proved that the set of \ $C^\infty$ \ Riemannian metrics 
 \ $\mathbf{g}$ \  on  \ $\mathbb{S}^2$ \ whose geodesic flow has \ $h_{top}(\mathbb{S}^2, \mathbf{g}) > 0$ \ is dense in the \ $C^2$ \ topology. Therefore, 
 a metric on \ $\Sigma$ \ is always close to a metric with positive topological entropy, and the former can be substituted by the latter for computational purposes,
 without changing drastically the system at hand, due to the continuity and differentiability properties of \ $h_{top}$ \ under such perturbations \cite{KKPW}.\\
 
 As mentioned above, determining the induced metric on \ $\Sigma$ \ depends on the field content of the theory, as well as on the particular foliation  
 of the horizon (the particular observer) employed in each case. In case that using the event horizon may not be desirable or feasible, as in the case of 
 numerical simulation of black holes, we can substitute the event horizon with marginally outer trapped surfaces \cite{CGP}. The stability of the time-like evolution 
 of the marginally outer trapped surfaces, an issue related to their possible topological type, was examined in \cite{AMS}.    \\
 
The last step is to establish a relation between the statistical entropy \ $\mathcal{S}_{BGS}$ \ of the black hole spacetime \ $(\mathcal{M}, \mathbf{g})$ \ and 
\ $h_{top}(\Sigma, \mathbf{g}_\Sigma)$. \ Before that, it may be worth recalling that \ $h_{top}(\Sigma, \mathbf{g}_\Sigma)$ \  characterizes the exponential 
growth rate of the number of distinguishable periodic orbits, with non-zero but arbitrary otherwise precision, of the geodesic flow of \ $\mathbf{g}$ \ on \ $\Sigma$. \ 
The choice of the topological entropy to measure the complexity of \ ($\Sigma, \mathbf{g}_\Sigma$) \ is desirable from another viewpoint too. 
Recall that, according to the variational principle \cite{KatHas}, 
\begin{equation}
       h_{top} (s) \ = \ \sup_{\mu} \  h_{KS}(\mu, s) 
\end{equation}
where \ $\mu$ \ ranges over all \ $s$-invariant probability measures on \ $\mathcal{M}$. \ In other words, when \ $h_{top} >0$, \ the variational principle implies that there
is an invariant measure \ $\mu$ \  whose  Kolmogorov-Sinai entropy \ $h_{KS}(\mu, s)$ \ is positive too. It is the Kolmogorov-Sinai entropy which is usually identified with the 
physical entropy in many dynamical  systems of physical interest. This is due to the fact that through Ruelle's inequality and Pesin's theorem, \ $h_{top}$ \ is related to the 
non-triviality of some of the Lyapunov exponents of 
the system \cite{KatHas} which quantify its chaotic behavior asymptotically, \ $s\rightarrow\infty$. \ 
Another advantage motivating the choice of topological entropy in the present context, is that it is usually easier to 
compute than the Kolmogorov-Sinai entropy \ $h_{KS}$.\\       

Following the dependence of \ $h_{top}$ \ on the linear dimensions of \ $\Sigma$, \  we see that the naive identification proposed 
in \cite{NK1}   
\begin{equation}   
      \mathcal{S}_{BGS} (\mathcal{M}, \mathbf{g}) \ \equiv \  \frac{1}{[h_{vol}(\Sigma, \mathbf{g}_\Sigma)]^2}
\end{equation}
is not valid in the present case. This may not be too surprising in view of the fact that there is no equality of the topological with the volume entropy 
for the geodesic flows on Riemannian manifolds \ ($\mathcal{M}, \mathbf{g})$ \  which have conjugate points.  
As a result and due to simplicity, one may be tempted to make the identification 
\begin{equation}
     \mathcal{S}_{BGS} (\mathcal{M}, \mathbf{g}) \ \equiv \ h_{top} (\Sigma, \mathbf{g}_\Sigma)
\end{equation}
maybe up to an overall sign, to make the signs in the mathematical and physical literature compatible. With this identification one may have to explain the transition
from the volume entropy to the topological entropy each  being the statistical entropy, in the degeneration from the genus \ $g$ \ surfaces \  $\Sigma$ \ to spheres
expressing the limit from semi-classical gravity  to  classical General Relativity. As this process  which was outlined in \cite{NK1} is highly conjectural, we will forego 
any further speculation how it may actually proceed and how its features may be  encoded through entropy. \\
  
A question that naturally arises is whether there is any relation between the embolic constant of a manifold  and the topological entropy of its geodesic flow, two 
quantities that are quite hard to explicitly calculate.
If such a relation existed, it would be an embolic analogue for the case of the 2-sphere, to 
 the Katz-Sabourau inequality \cite{KS}, which relates the systolic ratio to the volume entropy of a surface of genus \ $g$. \ In view of the Katz-Sabourau inequality, such a 
 relation might have some potentially interesting physical consequences. \\
   

\section{Conclusions and future outlook}

In this note, we presented a hand-waving argument purporting to show that the statistical entropy of a horizon in (3+1)-dimensional spacetime, 
whose sections \ $\Sigma$ \ have spherical topology, may be a function of the  topological entropy  of their geodesic flow, 
where the metric of \  $\Sigma$ \  is induced by the metric of the horizon. 
This complements the work of \cite{NK1} where the entropy of higher genus sections \ $\Sigma$ \ was presented. Here, as well as in \cite{NK1},
the results are curvature-free, something that may be advantageous, since lower bounds of the stress energy tensor in the Einstein equations are not known 
to hold for all spacetime vectors, or may not exist, even assuming the validity of the energy inequalities.\\

Potential extensions of the present work can be in various directions: the obvious extension is to higher dimensional black holes, ones that usually appear as a result of 
supergravity or of string/brane theories. This has been a topic of continuous interest for the last few decades \cite{Emp, Reall}.   
Another direction of potential interest is to investigate the microscopic origin of the black hole entropy and examine, in particular, 
whether it is of the usual Boltzmann/Gibbs/Shannon or of another, maybe power-law, functional form \cite{TC, CI1, NK2, CI2, Majhi}.  \\       

We believe that curvature-free results may be of wider interest in describing aspects of physical problems having a geometric or topological origin or description.
It might, for instance, be of some interest to explore if or to what extent any of the above concepts and methods may be beneficial in describing aspects of 
condensed matter systems such as topological insulators which are currently an area of considerable interest and intensive work \cite{Rachel, Shankar}.\\


               \vspace{10mm}

\noindent{\normalsize\bf Acknowledgement:}  \ \ We are  grateful to Professor Anastasios Bountis  for his encouragement and support without which this work
would have never been possible.  \\  


              \vspace{7mm}



\end{document}